\documentclass[10pt]{article}

\usepackage[margin=0.78in]{geometry}
\usepackage{amsmath,amssymb}
\usepackage{booktabs}
\usepackage{array}
\usepackage{microtype}
\usepackage{xcolor}
\usepackage{tikz}
\usetikzlibrary{arrows.meta,positioning,fit,calc}
\usepackage[numbers,sort&compress]{natbib}
\usepackage[hidelinks]{hyperref}
\usepackage{xurl}

\newcommand{\hash}[1]{{\small\path{#1}}}
\newcommand{\rb}{route block}

\newcommand{\wthirteen}{W13}
\newcommand{\wtwo}{W2}
\newcolumntype{L}[1]{>{\raggedright\arraybackslash}p{#1}}

\title{\textbf{Route-Block Membership Selects Packed-AWQ Arithmetic}\\
\large A Controlled Single-Fixture Mechanism Study}
\author{Lukas Stepanek\\\small\texttt{luki.step@proton.me}}
\date{July 2026}

\begin{document}
\maketitle

\begin{abstract}
Mixture-of-experts (MoE) inference first aligns routed tokens into padded
expert blocks, then executes packed quantized matrix multiplication over those
blocks. This preprocessing is often treated as bookkeeping. In one
pre-specified Qwen3-Coder AWQ layer-6 fixture on a pinned vLLM/Marlin build
and RTX~3090 runtime, we show that the tested route-block interventions select
exact packed arithmetic trajectories. Two fixed preconstruction histories
produced distinct native alignments and exact trajectories. Injecting the
opposite alignment transferred \wthirteen, activation, routed-\wtwo, and final
outputs. Permuting two routes within one block preserved each native
trajectory, while exchanging two prior-data-selected routes across the
boundary between expert-106 blocks 40 and 41 transferred the complete
opposite trajectory. Source- and binary-derived schedule geometry maps those
blocks to direct/full-K and split/global-reduction classes. Forcing a
single-slice 200-block grid made \wthirteen{} bitwise equal. Stable canonical
construction made both histories converge to a third exact trajectory. The
confirmatory cohort contains 70 valid cold processes and seven required
perturbation rejections. This is a causal mechanism result for one fixture,
not a prevalence, allocator, portability, or serving-impact claim.
\end{abstract}

\section{Introduction}

Packed weight-only quantization makes large-model inference practical by
reducing memory traffic while retaining GPU-friendly kernels
\cite{lin2023awq,frantar2024marlin}. In an MoE layer, however, the matrix
multiplication is preceded by dynamic routing and alignment: valid routed
token rows are grouped by expert and padded into fixed-size blocks. The
alignment's internal order is usually discussed as a dispatch or utilization
detail. Exact floating-point output also depends on which aligned block a
route enters because the downstream kernel may map neighboring block indices
to different reduction schedules.

This paper isolates that connection in a narrow controlled case. A fixed
preconstruction treatment selected one of two native atomic alignment orders
for the same inputs and packed weights. Existing evidence had already exposed
the affected routes, so our study is confirmatory rather than an independent
discovery cohort. Before the confirmatory runs, we pre-specified the histories,
route interventions, schedule intervention, exact trajectory predictions,
five cold-process replicates per cell/history, and rejection rules.

Our central abstraction is \emph{route-to-eight-row-block membership}. It
forgets order within a padded expert block but retains which valid flattened
route identifiers occupy each block. This abstraction connects dispatch to
the pinned Marlin schedule: expert-106 block 40 is assigned directly with a
full-K reduction, whereas block 41 is split across K and globally reduced.

The contribution is evidence, not breadth:

\begin{itemize}
  \item a precise \rb{} membership abstraction and a derivation from block
  index to reduction-schedule class for the pinned kernel;
  \item a 70-process prospective mechanism matrix in which cross-order and a
  two-route boundary intervention transfer all four measured arithmetic
  surfaces, while a within-block permutation preserves them;
  \item a schedule control in which grid 200 removes the split/global
  reduction and collapses the \wthirteen{} distinction; and
  \item a separately evaluated canonical construction that makes both
  histories converge to one third trajectory, with an auditable public
  recorded-evidence package.
\end{itemize}

We do not estimate how often the effect occurs. We do not identify an
allocator primitive, and we make no claim about model-token divergence,
serving impact, portability, paging, or prefetch behavior.

\section{Background}

\subsection{MoE alignment and padded blocks}

Let the top-$k$ router produce an $M\times k$ expert-ID array. Flattening in
row-major order gives route IDs $r\in[0,Mk)$. Alignment groups valid routes by
expert, pads each expert's count to a multiple of block size $b$, and emits a
parallel expert-block vector. The tested path uses $M=21$, $k=8$, 128 experts,
and $b=8$: 168 valid routes become 400 aligned rows in 50 active blocks.

MoE systems have long used blocking to reconcile dynamic routing with GPU
execution. MegaBlocks, for example, formulates sparse MoE training around
block-sparse operations to avoid token dropping and padding waste
\cite{gale2022megablocks}. Our concern differs: we hold the routed inputs and
weights fixed and ask whether block membership changes exact packed
arithmetic.

\subsection{Packed-AWQ Marlin arithmetic}

AWQ is an activation-aware weight-only quantization method
\cite{lin2023awq}. Marlin provides mixed-precision kernels with tiled
scheduling, pipelining, and parallel reduction \cite{frantar2024marlin}. The
tested vLLM path applies Marlin-style packed-AWQ MoE GEMMs to the aligned
rows. Floating-point addition is non-associative, so a route that moves
between a direct/full-K tile and a split-K tile with global reduction can
follow a different exact rounding trajectory even though its mathematical
dot product and packed weights are unchanged.

\section{The route-block abstraction}

\subsection{Definition}

For expert $e$, let $A_e$ be its aligned sequence after padding. Consecutive
groups of eight positions form expert-local blocks. We define
\begin{equation}
  B(e,j)=\{r:\ r\ \text{is valid for expert }e
  \text{ and occupies }A_e[8j:8j+8]\}.
\end{equation}
Padding sentinels are excluded and order inside $B(e,j)$ is ignored. The
complete mode identity is the ordered sequence of
$(e,j,\operatorname{sort}(B(e,j)))$. Consequently, swapping two valid routes
inside the same $B(e,j)$ preserves the abstraction, while moving a route
across an eight-row boundary changes it.

\begin{figure}[t]
\centering
\begin{tikzpicture}[
  node distance=4mm,
  route/.style={draw,rounded corners,minimum width=8mm,minimum height=6mm,font=\scriptsize},
  block/.style={draw,thick,rounded corners,inner sep=3mm},
  sched/.style={draw,rounded corners,align=center,minimum width=28mm,minimum height=11mm,font=\small},
  arrow/.style={-{Latex[length=2mm]},thick}
]
  \node[route] (r22) {22};
  \node[route,right=1mm of r22] (r28) {28};
  \node[route,right=1mm of r28] (r51) {51};
  \node[route,right=1mm of r51] (other40) {$\cdots$};
  \node[block,fit=(r22)(r28)(r51)(other40),
        label={[xshift=-10mm]below:{\scriptsize expert 106, block 40}}] (b40) {};
  \node[route,right=16mm of other40] (r86) {86};
  \node[route,right=1mm of r86] (other41) {$\cdots$};
  \node[block,fit=(r86)(other41),
        label={[xshift=-10mm]below:{\scriptsize expert 106, block 41}}] (b41) {};
  \node[sched,below=14mm of b40] (full) {direct\\full K};
  \node[sched,below=14mm of b41] (split) {split K\\global FP32 reduce};
  \draw[arrow] ([xshift=11mm]b40.south) -- ([xshift=11mm]full.north);
  \draw[arrow] ([xshift=11mm]b41.south) -- ([xshift=11mm]split.north);
  \draw[<->,very thick,red!70!black] (r51.north) to[bend left=28]
    node[above,font=\scriptsize] {boundary intervention} (r86.north);
  \draw[<->,blue!70!black] (r22.south) to[bend right=32] (r28.south);
  \node[font=\scriptsize,blue!70!black,align=center,left=7mm of b40]
    (withinlabel) {within-block\\control};
  \draw[-{Latex[length=1.5mm]},blue!70!black]
    (withinlabel.east) -- (r22.west);
\end{tikzpicture}
\caption{The tested instantiation of route-block membership. Routes 22 and 28
are permuted inside block 40; routes 51 and 86 are exchanged across the
40/41 schedule boundary. Other routes are omitted from the drawing.}
\label{fig:routeblock}
\end{figure}
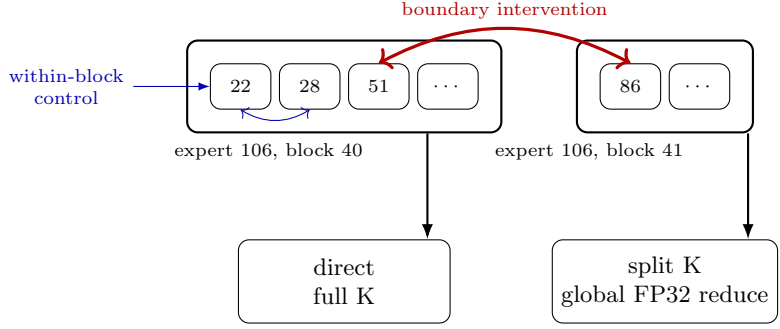

\subsection{Mapping block index to schedule class}

The captured public call has $N=1536$ and $K=2048$. The pinned auto-selector
chooses $\texttt{thread\_k}=64$, $\texttt{thread\_n}=128$, 128 threads, and
three resident blocks per SM. Each of the 50 MoE blocks therefore has 12
$N$-tiles and 32 $K$-tiles, for 600 global $(\text{MoE block},N\text{-tile})$
tiles.

With the native grid of 246 CUDA blocks, the first $2\cdot246=492$ global
tiles receive direct/full-K assignments. The remainder begins at MoE block
\begin{equation}
  \left\lfloor\frac{492}{12}\right\rfloor=41
\end{equation}
and is split into two 16-$K$-tile slices followed by global FP32 reduction.
Thus block 40 is direct/full-K and block 41 is split/reduced. Membership in
$B(106,40)$ versus $B(106,41)$ selects a reduction-schedule class
(Figure~\ref{fig:routeblock}).

At grid 200, the direct region covers 400 global tiles. The remainder begins
at block $\lfloor400/12\rfloor=33$ but receives one 32-$K$-tile slice. All
column tiles are therefore single-slice and the global reduction is bypassed.
This derivation was frozen before the grid-200 outcome. It corrects an earlier
transposition of \texttt{thread\_k} and \texttt{thread\_n}; the superseded
grid-240 prediction was rejected before arithmetic execution.

\section{Prospective study}

\subsection{Fixture, software build, and histories}

The fixture is layer 6 of the
\path{QuantTrio/Qwen3-Coder-30B-A3B-Instruct-AWQ} checkpoint at Hugging
Face revision
\hash{c58857a7f41c0920f73d1b56678640f9c02017d7}
\cite{quanttrio2025qwen3coder}. Its 17-file manifest has SHA-256
\hash{bf498217bf4da3d43864150212b630f08340d05e6922696dd7562539b11798e8};
the derived model identity is
\hash{94ab8051a1bfbb6671aebade458b55760ca2ab3432da057e385c4327ce4a286c}.
The layer has hidden size 2048 and uses 4-bit AWQ with group size 128. The
checkpoint-derived fixture has SHA-256
\hash{4f4c0812377ab927727a726ee94aafa453dc02933ebd6913c740a6d335d431a8}.
The software build pins vLLM commit
\hash{752a3a504485790a2e8491cacbb35c137339ad34}
\cite{vllm2026commit}; its tracked diff has SHA-256
\hash{1c366352593f896113785cb6a5e137ee2b646c267eb185a6decb3033f7239744}
and its final native extension has SHA-256
\hash{b941455ec4fcf8b4b567f51575271a7f7928922a2a8291d41abb9ed78b5165b5}.
The runtime is an RTX~3090 (compute capability 8.6), driver 580.159.03,
CUDA 13.0, and PyTorch 2.11.0+cu130. Every arm uses identical input,
top-$k$ IDs, top-$k$ weights, and resident packed weights.

H0 performs no extra allocation/prewarm prelude. Before model construction,
H1 allocates and retains three persistent CUDA tensors shaped like the
fixture's input, top-$k$ weights, and top-$k$ IDs; it copies the immutable
fixture into them and synchronizes once. They are never used as operator
inputs or outputs. H1 is a composite treatment: it may select a different
native alignment, but it cannot identify allocation, fragmentation, address
residue, prewarming, or a named allocator behavior as the cause.

\subsection{Pre-specified cells and analysis}

Routes 51/86 and 22/28 were selected from prior H0/H1 data. The former were
the only routes differing at the captured \wthirteen{} boundary; the latter
remained together in block 40. We disclosed this prior-data selection before
the new cross-order and surgical-intervention outcomes.

\begin{table}[t]
\centering
\small
\setlength{\tabcolsep}{4pt}
\begin{tabular}{@{}p{22mm}p{42mm}p{63mm}@{}}
\toprule
Cell & Intervention & Frozen exact prediction \\
\midrule
Native & untreated H0/H1 & distinct H0 and H1 trajectories \\
Own order & inject each history's own plan & reproduce its native trajectory \\
Cross order & inject the opposite plan & transfer all four opposite surfaces \\
Canonical & stable $(e,r)$ construction & both histories equal one third trajectory \\
Within block & permute routes 22/28 & preserve each native trajectory \\
Boundary & exchange routes 51/86 & transfer the complete opposite trajectory \\
Single slice & own plan, grid 200 & identical \wthirteen{}; no final-collapse prediction \\
\bottomrule
\end{tabular}
\caption{Pre-specified cells. Each cell/history has five independent
cold processes. Qualification contains 20 processes, causal transfer 20, and
ablation 30.}
\label{tab:cells}
\end{table}

The study executed replicate-major with history order alternated by
replicate. A separate scorer reloaded raw tensors, recomputed hashes and
\rb{} membership, enforced intervention and grid consumption, and required
five exact outcomes per cell/history. Seven mutations---route, expert block,
treatment marker, grid marker, stage hash, raw tensor, and source/binary
identity---were each rejected for the predeclared reason. Timing in this
cohort was operational telemetry and was not analyzed as performance.

\clearpage
\section{Results}

\subsection{Route-block membership transfers exact arithmetic}

All 70 prospective processes were valid. Table~\ref{tab:results} gives the
exact trajectory relation rather than an averaged metric. Native H0 and H1
were distinct at every captured surface. Injecting each native plan under its
own history reproduced its native trajectory, excluding intervention
machinery as the source of the difference. Cross-order injection transferred
all four surfaces to the opposite exact trajectory.

\begin{table}[h]
\centering
\small
\begin{tabular}{@{}lcccl@{}}
\toprule
Cell & H0 result & H1 result & Replicates & Interpretation \\
\midrule
Native & H0 & H1 & $5+5$ & distinct \\
Own order & H0 & H1 & $5+5$ & intervention-neutral \\
Cross order & H1 & H0 & $5+5$ & complete transfer \\
Canonical & C & C & $5+5$ & history collapse \\
Within block & H0 & H1 & $5+5$ & membership-preserving \\
Boundary & H1 & H0 & $5+5$ & two-route sufficiency \\
Grid 200 & S at W13 & S at W13 & $5+5$ & single-slice W13 collapse \\
\bottomrule
\end{tabular}
\caption{Exact results. H0, H1, and C are three distinct four-surface
trajectories. S is the shared grid-200 \wthirteen{} trajectory; activation,
\wtwo, and final grid-200 values were recorded but were not pass criteria.}
\label{tab:results}
\end{table}

Changing only the order of routes 22 and 28 within block 40 preserved each
history's native trajectory. Exchanging routes 51 and 86 across blocks 40 and
41 transferred the complete opposite trajectory, although the other 18
H0/H1 membership changes remained untouched. In this fixture, those two
prior-data-selected membership changes were sufficient for the observed
transfer.

The native difference was sparse: seven FP16 \wthirteen{} elements on routes
51 and 86 (maximum absolute difference $7.62939453125\times10^{-6}$), three
activation elements on rows 17 and 86
($9.5367431640625\times10^{-7}$), ten routed-\wtwo{} elements on route 51
($7.62939453125\times10^{-6}$), and one final element on row 6
($9.5367431640625\times10^{-7}$). These counts describe this fixture; they
are not prevalence or practical-impact estimates.

\subsection{The schedule control}

Under the native 246-block grid, block 40 and block 41 occupy the two derived
schedule classes. At grid 200, H0 and H1 \wthirteen{} tensors became bitwise
equal with shared SHA-256 prefix \hash{44118553b1d46f3f}. This confirms the
pre-specified schedule-class explanation at \wthirteen{}. It does
not isolate one instruction or one FP32 addition, and it does not imply final
output collapse.

\subsection{Canonical third trajectory}

Stable construction orders valid routes by $(\text{expert ID},
\text{flattened route ID})$. Under both histories it produced the same
\rb{} fingerprint and the same four surface hashes. Its final hash begins
\hash{717d4f34}, whereas native H0 and H1 begin \hash{bdfadfc9} and
\hash{1c317016}; canonical construction is a third exact trajectory, not
recovery of either historical native trajectory.

\begin{figure}[h]
\centering
\begin{tikzpicture}[
  state/.style={draw,rounded corners,minimum width=24mm,minimum height=9mm,font=\small},
  lab/.style={font=\scriptsize,align=center},
  arrow/.style={-{Latex[length=2mm]},thick}
]
  \node[state] (h0) {native H0};
  \node[state,right=46mm of h0] (h1) {native H1};
  \node[state,below=22mm of $(h0)!0.5!(h1)$] (c) {canonical C};
  \draw[arrow,bend left=15] (h0) to node[above,lab]{opposite order or\\51/86 boundary swap} (h1);
  \draw[arrow,bend left=15] (h1) to node[below,lab]{opposite order or\\51/86 boundary swap} (h0);
  \draw[arrow] (h0) -- node[left,lab]{stable\\construction} (c);
  \draw[arrow] (h1) -- node[right,lab]{stable\\construction} (c);
\end{tikzpicture}
\caption{Exact trajectory relations. The canonical result is a third
trajectory shared by both histories; it is not restoration of H0 or H1.}
\label{fig:trajectories}
\end{figure}
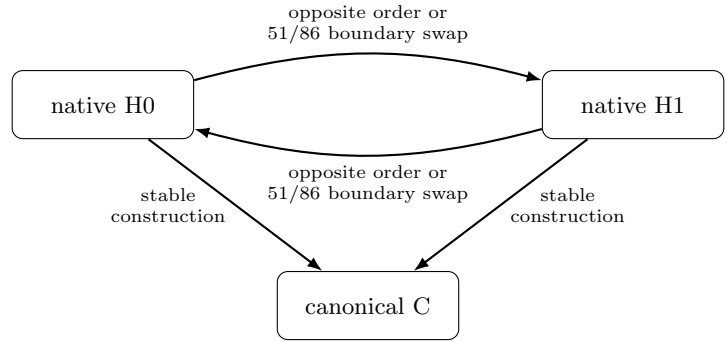

\section{Separate evaluation of canonical construction}

The canonical implementation was evaluated in a separate, pre-specified
cohort on a minimal eight-file vLLM patch. It uses one CUDA thread
per expert to scan flattened routes in ascending route ID, is disabled by
default, and is scoped to AutoAWQ-Marlin. The evaluation recorded 104
randomized alignment cases against an exact CPU reference, 1,024 fixed-vector
packed-AWQ replays, all 48 MoE layers (480 observations in 10 processes), and
30 hook-free cold processes across code, chat, and long-context prompts.
Unsupported configurations failed closed.

The mechanism-study and implementation-evaluation builds share the exact vLLM
base, model identity, model manifest, layer-6 fixture, GPU/runtime
anchors, and seven byte-identical deterministic operator files. They have
different complete diffs and native extension hashes and were executed as
different populations. The separate cohort therefore corroborates the
implementation solution; its processes are not added to the prospective 70.

\subsection{Exact single-operating-point measurements}

A separate pre-specified performance protocol fixed the software build,
operating points, warmups, timing boundaries, statistics, and non-inferiority
margins. Its independent validator recomputed the medians and ratios, so we
report them once and without a general performance label.

\begin{table}[h]
\centering
\small
\begin{tabular}{@{}lrrr@{}}
\toprule
Operating point & Canonical & Stock & Canonical/stock \\
\midrule
Layer-6 call (ms) & 0.2412 & 0.2416 & 0.9986 \\
Batch-1 fixed prompt (tokens/s) & 28.401 & 28.362 & 1.0014 \\
\bottomrule
\end{tabular}
\caption{Two single-operating-point medians from the separate pre-specified
implementation evaluation; full-precision values remain in the artifact.
Layer timing used seven alternating
50-call CUDA-event blocks after 20 warmups per mode. End-to-end timing used
five alternating requests, one warmup per mode, and 64 forced greedy tokens.
These data do not establish negligible overhead, speedup, serving-scale
throughput, or portability.}
\label{tab:performance}
\end{table}

\section{Historical corroboration}

A prior stock-versus-4-MiB-candidate comparison showed bidirectional order transfer,
canonical collapse to the same \hash{717d4f34...} output, and grid-200
\wthirteen{} collapse. That contrast was construction-confounded: 36 routes
changed block membership and its highlighted routes were 22, 28, 109, and
126. It is corroborating evidence only. We neither relabel its arms as H0/H1
nor count its runs in the 70-process cohort.

Separate historical controls tested whether initialized/preallocated state
alone restored the historical stock trajectory. It did not: H0 matched the
current-candidate result in five of five runs, while H1 matched the historical
candidate in four of five and stock in one of five. This rules out reliable
restoration of the stock trajectory but does not establish broad allocator
causality.

\section{Related work}

Table~\ref{tab:related} contrasts our route-block mechanism with AWQ
quantization \cite{lin2023awq}, Marlin scheduling \cite{frantar2024marlin},
MegaBlocks block-sparse MoE training \cite{gale2022megablocks}, UniEP
deterministic token ordering \cite{zheng2026uniep}, and Bit-Exact inference
invariance \cite{cankaya2026bitexact}. WiSP reports output-preserving vLLM
expert paging but leaves GPTQ/AWQ to future work and does not study
alignment-induced packed arithmetic \cite{zhang2026wisp}.

Two contemporaneous, open vLLM software records are especially adjacent.
PR~46639 proposes full-$K$ Marlin execution to remove batch-size-dependent
split-reduction effects \cite{kouba2026batchinvariance}; PR~48032 proposes
stable route alignment after tracing Marlin output differences to atomic
alignment order \cite{brown2026routealignment}. Neither record reports the
selected eight-row boundary intervention, within-block control,
schedule-collapse control, and canonical third trajectory tested here. Both
were unmerged when accessed and are cited as public engineering records, not
peer-reviewed results.

\begin{table}[t]
\centering
\scriptsize
\setlength{\tabcolsep}{3pt}
\begin{tabular}{@{}L{19mm}L{30mm}L{29mm}L{48mm}@{}}
\toprule
Work & Primary focus & Ordering/determinism & Relation to this study \\
\midrule
AWQ & weight-only quantization & not the focus & supplies the quantized model form \\
Marlin & packed mixed-precision kernels & complex tiled scheduling & supplies the pinned schedule whose classes we derive \\
MegaBlocks & block-sparse MoE training & block organization for efficiency & related blocking, not exact packed-arithmetic causality \\
UniEP & expert-parallel training megakernel & deterministic token order & related canonical ordering under a different system and workload \\
Bit-Exact & inference verification & determinism versus invariance & related exact-output framing, not the route-block intervention \\
vLLM PR 46639 & WNA16 batch invariance & full-$K$ versus split-$K$ & adjacent schedule control; no selected boundary-transfer study \\
vLLM PR 48032 & Marlin route alignment & stable route ordering & adjacent order control; no schedule-class causal matrix \\
This work & single-fixture mechanism & route-block membership and stable construction & causal boundary, within-block, schedule, and canonical controls \\
\bottomrule
\end{tabular}
\caption{Compact comparison with the closest conceptual lines of work.}
\label{tab:related}
\end{table}

\section{Limitations and threats to validity}

\paragraph{One selected fixture.}
The study has one independent problem instance: one checkpoint, layer,
prompt-derived fixture, GPU, runtime, and kernel build. Five cold processes
per cell establish repeatability, not external validity. The focal routes
were selected from prior data. We do not estimate a denominator or
prevalence.

\paragraph{Composite treatment.}
H1 is a fixed preconstruction-state treatment. It establishes association
between that composite treatment and native alignment, followed by causal
control of alignment and arithmetic. It does not identify allocator,
fragmentation, address, or prewarm primitives.

\paragraph{Pinned arithmetic explanation.}
The source/binary derivation and grid control establish direct/full-K versus
split/global-reduction schedule classes for one pinned Marlin geometry. They
do not identify a single responsible instruction or generalize across GPU
architectures, compiler versions, shapes, or kernels.

\paragraph{Exact differences are not practical impact.}
The final native difference is one FP16 element. We did not test token
divergence, task quality, or serving impact. Sparse exact differences can be
scientifically diagnostic without being operationally important.

\paragraph{Separate implementation evidence.}
The canonical-implementation evaluation shares exact anchors but uses a
different full diff, binary, and process population. Its single-point timings
cannot support general performance, serving-scale, or portability claims.

\paragraph{Software-build linkage.}
The corrected schedule geometry was derived with a qualified precursor
extension, whereas the 70-process cohort used the final extension identified
above. The binaries differ. The pinned \texttt{ops.cu} and
\texttt{marlin\_template.h} schedule sources are byte-identical across the
two builds; the artifact records both source hashes and both extension
identities. This source join supports the schedule derivation but does not
combine the two process populations.

\paragraph{Recorded versus native reproducibility.}
The public artifact includes the exact pre-specified plans, capture/scorer sources,
software-build diff, pre-execution study record, and execution records. It omits
checkpoint-derived tensors, private prompt, binary extensions, and raw tensor
captures. Its verifier checks recorded evidence and exact relations. Exact
independent reproduction of the reported bit patterns is unavailable from the
public package because the omitted prompt and fixture tensors are required to
reconstruct the reported fixture; rebuilding the pinned environment with a new
fixture is a methodological replication, not confirmation of those bit
patterns.

\paragraph{Publication chronology.}
The contracts and outcomes were first published together after execution.
The companion artifact exposes byte-level cross-linkage and recorded
execution chronology, not an independent third-party preregistration
timestamp.

\section{Artifact and evidence discipline}

The companion artifact is hosted at
\url{https://github.com/gustavgauge/route-block-arithmetic}; the immutable
\texttt{arxiv-v1-submission-r2} release is the paper's submission package and
links to the unchanged recorded-evidence snapshot.
The exact evidence snapshot is Git commit
\hash{f17755bf90dbca97ddee78ad70d1bfc58ad282e2}. Its sole
claim/evidence/provenance ledger separates five classes: the prospective
cohort, schedule derivation, historical corroboration and controls, the
canonical-implementation evaluation, and unsupported claims. Frozen
contracts, stage summaries and independent scores, perturbation record,
schedule geometry, implementation verdict, exact timing record, historical
inventory, and checkpoint manifest are SHA-256 pinned. The CPU-only verifier
recounts the 20/20/30 stage populations and all cell/history replicates;
verifies trajectory equalities and transfers; checks the schedule arithmetic;
confirms all seven rejection records; validates the historical/cohort
separation and checkpoint identity; verifies the pre-execution study record,
software-build identities, plans, execution records, and scorer sources; and
recomputes the separate timing medians
and ratios. It is intentionally not a GPU tensor re-execution.

\paragraph{Use of AI tools.}
The author used OpenAI Codex for code execution, evidence cross-checking, and
drafting and editing assistance. The author reviewed the analyses, verified
the cited artifacts, and assumes responsibility for the paper's contents.

\section{Conclusion}

In this controlled fixture, MoE alignment acted as arithmetic control flow.
The tested route-block boundary intervention changed which packed reduction
schedule the selected routed rows entered and transferred their exact
trajectory. The within-block permutation preserved each native trajectory,
removing split reduction collapsed \wthirteen, and stable construction removed
the observed history dependence by selecting a third exact trajectory. The
evidence supports that mechanism at the stated boundary---no broader claim is
required for the result to be useful or falsifiable.

\bibliographystyle{plainnat}
\bibliography{references}

\end{document}